\documentclass{PoS}

\title{Thermal dilepton production from hot QCD}

\ShortTitle{Thermal dilepton production from hot QCD}

\author{M.\ Laine%\speaker{Mikko Laine}%
         \thanks{Supported in part by
         the Swiss National Science Foundation
         (SNF) under grant 200020-155935.}\\
        Institute for Theoretical Physics, AEC, University of Bern, 
        Sidlerstrasse 5, 3012 Bern, Switzerland\\
        E-mail: \email{laine@itp.unibe.ch}}

\abstract{%
NLO and LPM-resummed computations of thermal dilepton production 
from a hot QCD plasma are reviewed. An interpolation applicable for 
all positive photon invariant masses is presented, and possibilities 
for comparisons with lattice and experimental data are pointed out.
}

\FullConference{9th International Workshop on Critical Point and Onset of
Deconfinement\\
                 17-21 November, 2014\\
                 ZiF (Center of Interdisciplinary Research), University of
Bielefeld, Germany}

\begin{document}

% put your own definitions here:
\newcommand{\ko}{k_0}
\newcommand{\km}{k_-}
\newcommand{\kp}{k_+}
\newcommand{\nB}[1]{n_\rmii{B{#1}}}
\newcommand{\nF}[1]{n_\rmii{F{#1}}}
\newcommand{\rmO}{{\mathcal{O}}}
\newcommand{\mE}{m_\rmi{E}}
\newcommand{\fe}{\rmi{f}}
\newcommand{\bo}{\rmi{b}}
\newcommand{\eq}{eq.~}
\newcommand{\eqs}{eqs.~}
\newcommand{\Tint}[1]{{\hbox{$\sum$}\!\!\!\!\!\!\!\int}_{\!\!\!\!#1}}
\renewcommand{\vec}[1]{{\bf #1}}
\newcommand{\nwc}{\newcommand}
\newcommand{\pf}{\frac{1}{16\pi^2}}
\newcommand{\bfx}{{\bf x}}
\newcommand{\bfi}{{\bf i}}
\newcommand{\tinymsbar}{{\overline{\mbox{\tiny\rm{MS}}}}}
\nwc{\nl}  {\newline}
\nwc{\be}  {\begin{equation}}
\nwc{\ee}  {\end{equation}}
\nwc{\bmu} {\bar{\mu}}
\nwc{\ba}  {\begin{eqnarray}}
\nwc{\ea}  {\end{eqnarray}}
\nwc{\bc}  {\begin{center}}
\nwc{\ec}  {\end{center}}
\nwc{\bi}  {\begin{itemize}}
\nwc{\ei}  {\end{itemize}}
\nwc{\nn}  {\nonumber\\}
\nwc{\Tr}  {\mathop{\rm Tr}}
\nwc{\re}  {\mathop{\rm Re}}
\nwc{\im}  {\mathop{\rm Im}}
\nwc{\Hc}  {\mathop{\rm H.c.}}
\newcommand{\aL}{a^{ }_\rmiii{L}}
\newcommand{\aR}{a^{ }_\rmiii{R}}
\newcommand{\gammaE}{\gamma_\rmiii{E}}
\newcommand{\rmiii}[1]{{#1}} % {{\mbox{\tiny{$\scriptstyle{\rm#1}$}}}}
\nwc{\la}[1]{\label{#1}}
\nwc{\rmi}[1]{{\mbox{\scriptsize #1}}}
\nwc{\nr}[1]{(\ref{#1})}
\nwc{\fr}[2]{{\frac{#1}{#2}}}
\nwc{\msbar}{\overline{\mbox{\rm MS}}}
\nwc{\lambdamsbar}{\Lambda_{\overline{\rm MS}}}
\nwc{\dr}{{4d\to3d}}
\newcommand{\Nf}{N_{\rm f}}
\newcommand{\Nc}{N_{\rm c}}
\newcommand{\Tc}{T_{\rm c}}
\newcommand{\mention}[2]{\hfill\parbox[c]{#1}{\tiny \ts \hfill #2}}
\newcommand{\rmii}[1]{{\mbox{\tiny\rm{#1}}}}
\newcommand{\mD}{m_\rmi{D}}
\newcommand{\ir}{\rmii{UV}} %{\rmii{IR}}
\newcommand{\Cf}{C_{\rm F}}
\newcommand{\CF}{C_\rmii{F}}
\newcommand{\betaL}{\beta_{\rmi{L}}}
\newcommand{\Sigmavi}{\Sigma_{v^{-1}}}
\newcommand{\gB}{g_\rmii{B}}
\newcommand{\fig}{fig.~}
\newcommand{\alphas}{\alpha_\rmi{s}}

\def\slash#1{#1\!\!\!/\!\,\,} 
\def\bslash#1{#1\!\!\!\!/\,\,} 
\def\lsi{\raise0.3ex\hbox{$<$\kern-0.75em\raise-1.1ex\hbox{$\sim$}}}
\def\gsi{\raise0.3ex\hbox{$>$\kern-0.75em\raise-1.1ex\hbox{$\sim$}}}
\nwc{\lsim}{\mathop{\lsi}}
\nwc{\gsim}{\mathop{\gsi}}

\newcommand{\rr}{{\mathbb{R}}}
\newcommand{\zz}{{\mathbb{Z}}}
\newcommand{\qq}{{\mathbb{Q}}}
\newcommand{\cc}{{\mathbb{C}}}
\newcommand{\unit}{{\mathbbm{1}}} %{\ii}

%%%%%%%%%%%%%%%%%%%%%%%%%%%%%% SECTION %%%%%%%%%%%%%%%%%%%%%%%%%%%%%%%%%%
%
\section{Introduction and observables}
\label{se:intro}

The thermal production rate of off-shell photons, subsequently decaying
into on-shell lepton--antilepton pairs, is a rich source
of information concerning the properties of the hot QCD medium generated in 
heavy ion collision experiments. If invariant masses corresponding to 
known vacuum resonances are avoided, the result can be expected to 
be relatively free from non-thermal background effects. Conversely, 
a focus on thermal modifications of prominent resonances, 
particularly quarkonium states, can in itself be 
turned into a useful probe of medium properties. 
In the present contribution we 
concentrate on non-resonant production and in particular on 
the contribution from gluons and three ($\Nf = 3$)
deconfined light quark flavours. 

In order to establish notation, let us denote by $T$ the temperature; 
by $k \equiv |\vec{k}|$ the total momentum of the dilepton pair  
with respect to the plasma rest frame; by $\ko$ the pair's energy; 
and by  
\be
 M \equiv \sqrt{\ko^2 - k^2}   
\ee 
its invariant mass.
To leading order in the electromagnetic fine-structure constant $\alpha_e$
and for massless quarks, the differential 
production rate per volume reads~\cite{old1,old2,old3}
\be
 \frac{{\rm d} \Gamma_{\mu^-\mu^+}}
   {{\rm d}\ko {\rm d}^3\vec{k}} 
   \stackrel{M^2 \ll\, m_Z^2 }{=}    
 - \frac{ \nB{} (\ko)  } 
  {3 \pi^3 M^2} 
 \, \theta(M^2 - 4 m_\mu^2)
 \, \biggl( 
   1 + \frac{2 m_\mu^2}{M^2}
 \biggr)
 \, \biggl(
   1 - \frac{4 m_\mu^2}{M^2} 
 \biggr)^\fr12 
 \, \alpha_{e}^2 
 \sum_{i = 1}^{\Nf} Q_i^2 \;  
 \im \Pi^{ }_\rmii{R}(\ko,k)
 \;. \la{dilepton_rate}
\ee 
Here $\nB{}$ is the Bose distribution, $Q_i$ the electric charge of 
a quark of flavour $i$ in units of the electron charge,  
and $\im \Pi^{ }_\rmii{R}$ stands for the imaginary part of a retarded
correlator (i.e.\ a spectral function) of one flavour, 
evaluated in an equilibrium 
ensemble at a temperature $T$.  

Given that we are interested in 
making contact with lattice simulations, the retarded correlator
is best interpreted as an analytic continuation of  
an imaginary-time correlator (cf.\ e.g.\ ref.~\cite{analytic}). 
Denoting the imaginary-time  coordinates by
\be
 X \equiv (\tau,\vec{x}) \;, \quad
 K \equiv (\omega^{ }_n,\vec{k}) \;, 
\ee
with $\omega_n \equiv 2\pi T n$, $n \in \zz$, 
the analytic continuation reads
\be
 \Pi^{ }_\rmii{R}(\ko,k)
  \equiv  
  \int_0^{1/T} \! {\rm d}\tau  
  \; e^{\,i \omega^{ }_n\tau }
  \int^{ }_\vec{x} 
   e^{\, i\, \vec{k} \cdot \vec{x}}
  \left\langle
    {{J}}^\mu ({X}) \;
    {{J}}^{ }_{\mu}(0)
  \right\rangle^{ }_{\omega^{ }_n \to -i [\ko + i 0^+]}
  \;,
  \quad 
  {{J}}^\mu 
   \equiv  \bar{\psi} \gamma^\mu \psi
  \;, \la{def2}
\ee
where the spinor $\psi$ represents one flavour,  
and a suitable ultraviolet regularization is needed for defining
the Fourier transform at short separations. 
It is interesting to also consider the space-like correlator
\be
 V^{ }_{\mu\nu}(\omega^{ }_n,z) \equiv
 \int_0^{1/T} \! {\rm d}\tau \; e^{\,i \omega^{ }_n\tau }
 \int \! {\rm d}^{2} \vec{x}^{ }_\perp
 \, 
  \left\langle
    {{J}}^{ }_\mu (X)
    \; {{J}}^{ }_{\nu}(0)
  \right\rangle^{ }_{ }
 \;, \quad
 \vec{x} \, \equiv \, (\vec{x}^{ }_\perp,z)
 \;. 
\ee
Clearly, 
\be
 \Pi^{ }_\rmii{R}(\ko,k) = \biggl\{ 
 \int_{-\infty}^{\infty} \! {\rm d}z \; e^{i k z} 
 \; {V^{\mu}}_\mu(\omega^{ }_n,z)  
 \biggr\}^{ }_{\omega^{ }_n \to -i [\ko + i 0^+]}  
 \;.  \la{rela2}
\ee
The last equation suggests
that spatial correlations measured with $\omega^{ }_n \neq 0$ have
a principal relation to the real-time dilepton production rate
captured by $\im\Pi^{ }_\rmii{R}$.

%%%%%%%%%%%%%%%%%%%%%%%%%%%%%% SECTION %%%%%%%%%%%%%%%%%%%%%%%%%%%%%%%%%%
%
\section{Different regimes and previous work}
\label{se:regimes}

If \eq\nr{def2} is addressed within the {\em weak-coupling expansion}, 
the method to be used depends on the parametric magnitudes of $k$
and $M$. In the following we recall the main cases, denoting by 
$g \equiv \sqrt{4\pi\alphas}$ the gauge coupling.  For reference, 
the leading-order result reads
\be
  - \im \Pi_\rmii{R}  =  \frac{ \Nc T M^2 }{2\pi k }
 \ln\left\{
   \frac{\cosh\bigl(\frac{\kp}{2 T} \bigr) }
        {\cosh\bigl(\frac{\km}{2 T} \bigr) }  
 \right\}
 \;, \quad
 k^{ }_\pm \; \equiv \; \frac{\ko \pm k}{2}
 \;. \la{lo}
\ee

The simplest case to discuss is $k=0$, because then the result only 
depends on a single kinematic variable, $M = \ko$. If $M\gsim \pi T$, 
then the NLO correction to \eq\nr{lo} is infrared finite 
and small~\cite{nlo1,nlo2,nlo3} (these results have recently
been extended to a finite quark mass~\cite{yb}). 
However, the NLO correction increases
rapidly as $M$ decreases; for $M\sim gT$, the correction is 
of $O(1)$ and needs to be summed to all orders, yielding
a large enhancement~\cite{htl1}. More recently, it has been realized
that the original (``HTL'') resummation is not sufficient 
for capturing all relevant effects for $M \lsim gT$. 
The correct infrared behaviour, including a transport peak in 
$\im\Pi^{ }_\rmii{R} / \ko$ whose width is $\sim g^4 T/\pi^3$ 
and whose height determines the electric conductivity $\sim \alpha_e T/g^4$, 
has been worked out in numerical form in ref.~\cite{lpm3}.

The phenomenologically perhaps most interesting case concerns  
the production rate of dilepton pairs
with a ``soft'' invariant mass ($M \sim gT$)
but large spatial momentum ($k \gsim \pi T$). In this regime
the NLO-rate has a logarithmic singularity, which is regulated
(as indicated below by $M \to g T$)
by Landau damping of the spacelike quarks mediating $t$-channel 
exchange~\cite{htl2,htl3}:
\be
 \quad
  - \im \Pi_\rmii{R} = ... +  
  \frac{\alphas \Nc \CF T^2}{2} 
  \ln \left( \frac{T}{M\to g T} \right)
  \Bigl[ 1 - 2 \nF{}(k) \Bigr]
 \;, \la{log}
\ee
where $\nF{}$ is a Fermi distribution. 
In addition, there are finite terms which all contribute 
at the same order because of collinear enhancement, 
and need to be handled through Landau-Pomeranchuk-Migdal (LPM) 
resummation~\cite{lpm1} (LPM resummation incorporates HTL resummation
in an approximation valid for $k\gg g T$).
In order to avoid double counting, LPM resummation 
needs to be carefully combined with other processes~\cite{lpm2}.

In a ``hard'' regime $M \gg \pi T$, 
{\em Operator Product Expansion} (OPE) 
techniques become applicable~\cite{ope}. 
The result is available in a closed form up to NLO:
\be
 - \im\Pi^{ }_\rmii{R} 
 = \frac{\Nc {M}^2}{4\pi} 
 \left(\! 1 + \frac{3 \alphas \CF}{4\pi} \!\right)  
 + 
 \frac{4 \alphas \Nc \CF}{9}
 \left(\! 1 + \frac{4 k^2}{3 M^2} \!\right)
 \frac{\pi^2 T^4}{M^2}
 + \rmO\Bigl( \frac{\alphas T^6}{M^4} \Bigr)
 \;. \la{ope}
\ee 

Yet another approach is to carry out {\em lattice simulations}. 
Lattice QCD
being formulated in imaginary time, with a time coordinate
$0 < \tau < 1/T$, it is however not possible to measure the rate directly, 
but rather a particular transformation thereof: 
\be
 G_\rmii{E}(\tau,{k}) = 
 \int_0^\infty \! \frac{{\rm d}\ko}{\pi} \, \im \Pi^{ }_\rmii{R} (\ko,{k}) 
 \frac{\cosh\bigl( \frac{1}{2T} - \tau \bigr) \ko }
 {\sinh\bigl(\frac{\ko}{2T} \bigr)}
 \;. \la{relation}
\ee
This means that a part of the contribution comes from the ``unphysical'' 
domain $\ko < k$. Even though it is in principle
possible to invert the relation in \eq\nr{relation}
for $\im \Pi^{ }_\rmii{R} (\ko,{k})$~\cite{analytic},
in practice large systematic uncertainties are induced~\cite{cond}. 
A more controlled approach is to insert
an analytically determined $\im \Pi^{ }_\rmii{R}(\ko,{k})$ 
into \eq\nr{relation}
and compare the resulting $G_\rmii{E}(\tau,{k})$ directly with numerical 
measurements (for recent work 
and references, see ref.~\cite{latnew}).

%%%%%%%%%%%%%%%%%%%%%%%%%%%%%% SECTION %%%%%%%%%%%%%%%%%%%%%%%%%%%%%%%%%%
%
\section{Methods for recent developments}
\label{se:technical}

We now turn to works whose scope can be summarized as follows: 
\begin{itemize}

\item[(i)]
The rate $\im\Pi^{ }_\rmii{R}$ has been determined up to NLO 
in a ``generic'' regime $k, M \sim \pi T$, 
verifying the cancellation of infrared divergences and finding 
in general a small correction~\cite{master,dilepton}.

\item[(ii)]
For a hard momentum $k \sim \pi T$
there is an additional parametric 
scale, $M \sim g^{1/2} T$, across which NLO results valid
for $M \gsim \pi T$ and LPM-resummed results valid for $M \lsim gT$ 
can be interpolated into each other~\cite{lpm}.

\item[(iii)]
It has been suggested that the relation in \eq\nr{rela2} can 
be turned into a non-trivial crosscheck of dilepton
production rate computations, in the sense that the screening masses
associated with the $|z| \gg 1/T$ behaviour $V^{ }_{\mu\nu}(\omega^{ }_n,z)$
can be measured non-perturbatively on one hand, and computed through
an LPM-type resummation on the other. This permits for a direct
comparison of the two approaches, 
without analytic continuation~\cite{screening}. 
It has also been speculated, and demonstrated
within the AdS/CFT framework, that a direct ``analytic 
continuation'' of screening masses (rather than correlation
functions) might allow for a determination
of the electric conductivity~\cite{screening2}.

\item[(iv)]
Finally, in an impressive computation, 
the LPM-resummed analysis of ref.~\cite{lpm1} has 
been extended up to NLO in the soft regime $M \lsim gT$, 
verifying the predictions' stability~\cite{gm}.

\end{itemize}

The methods used in these studies are as follows. 
For (i), standard (unresummed) perturbation theory is sufficient, 
with the only complication that because of infrared divergences appearing
in ``real'' and ``virtual'' corrections, intermediate stages of the results
need to be worked out in the presence of an infrared regulator, which
cancels in the end~\cite{master,dilepton}.

For (ii), an essential ingredient is to realize that the 
result of the soft regime consists of two parts, the LPM-resummed
one and another part, referred to as $2\to 2$ scatterings, 
which also appears in the non-resummed NLO 
expression.  So, for a consistent result in the soft regime, the LPM-resummed
result and the NLO result need to be summed together, but only after 
subtracting those terms from the NLO result which got resummed
into the LPM one. This can be summarized as
\be
 \left. \im \Pi^{ }_\rmii{R} \right|^{ }_\rmi{interpolant} \; \equiv \;
 \left. \im \Pi^{ }_\rmii{R} \right|^{ }_\rmii{NLO}
 \; 
 - \; \left. \im \Pi^{ }_\rmii{R} \right|^\rmi{expanded to NLO}_\rmii{LPM}
 \; + \; \left. \im \Pi^{ }_\rmii{R} \right|^\rmi{full}_\rmii{LPM}
 \; 
 \;. \la{match}
\ee
This expression is correct in the soft regime $M \lsim gT$ 
because the subtraction
removes the danger of double-counting terms appearing
in 
$
 \left. \im \Pi^{ }_\rmii{R} \right|^{ }_\rmii{NLO}
$, but 
also in the hard regime $M \gsim \pi T$, 
because the two variants of the LPM expression cancel against
each other there (up to higher-order corrections). 

%%%%%%%%%%%%%%%%%%%%%%%%%%%% FIGURE %%%%%%%%%%%%%%%%%%%%%%%%%%%%%%%%%%%%%%%%%
%
\begin{figure}[t]

 \centerline{
    \includegraphics[angle=0,width=8.0cm]{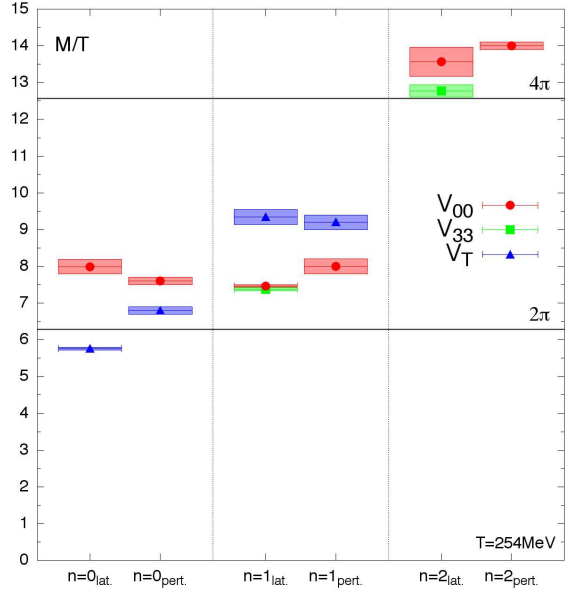}
 }

 \caption[a]{A comparison of screening masses related to different
 polarization states of the  
 vector current in the presence of a Matsubara frequency
 ($\omega_n = 2\pi T n$) in dynamical $\Nf = 2$ QCD at 
 $T\sim 250$~MeV (from ref.~\cite{screening}). 
 The correlation functions were measured in 
 the direction of the $z$-axis, and were averaged over the 
 ``transverse'' $(x^{ }_1,x^{ }_2)$ plane. 
 The good agreement between lattice
 measurements (``$\mbox{ }^{ }_\rmi{lat}$'') and resummed perturbative
 results (``$\mbox{ }^{ }_\rmi{pert}$'') suggests that the systematic
 uncertainties of both approaches are moderate.
 }
 \label{fig:screening}
\end{figure}
%
%%%%%%%%%%%%%%%%%%%%%%%%%%%%%%%%%%%%%%%%%%%%%%%%%%%%%%%%%%%%%%%%%%%%%%%%%%%%

For (iii), two ingredients are needed. One is a theoretical analysis 
showing that a perturbative determination of the screening masses at
$\omega^{ }_n \neq 0$ involves solving inhomogeneous Schr\"odinger-type
equations which are just analytic continuations of the corresponding
equations appearing in LPM resummation (with, in particular, the 
same ``potential''~\cite{sch2,mp} 
and the same ``inhomogeneous terms''). The other
ingredient is measuring screening masses with standard lattice QCD; 
in ref.~\cite{screening} this was done for $\Nf = 2$ 
light dynamical flavours. 

For (iv), the parameters appearing in LPM resummation,
in particular the asymptotic thermal masses and the potential, 
need to be modified through NLO corrections. Many ingredients
are known from previous work~\cite{sch2,sch1,ak}, but need to be 
put together in a consistent fashion~\cite{gm}.

%%%%%%%%%%%%%%%%%%%%%%%%%%%%%% SECTION %%%%%%%%%%%%%%%%%%%%%%%%%%%%%%%%%%
%
\section{Results}
\label{se:comparison}

%%%%%%%%%%%%%%%%%%%%%%%%%%%% FIGURE %%%%%%%%%%%%%%%%%%%%%%%%%%%%%%%%%%%%%%%%%
%
\begin{figure}[t]

\centerline{
    \includegraphics[angle=0,width=8.0cm]{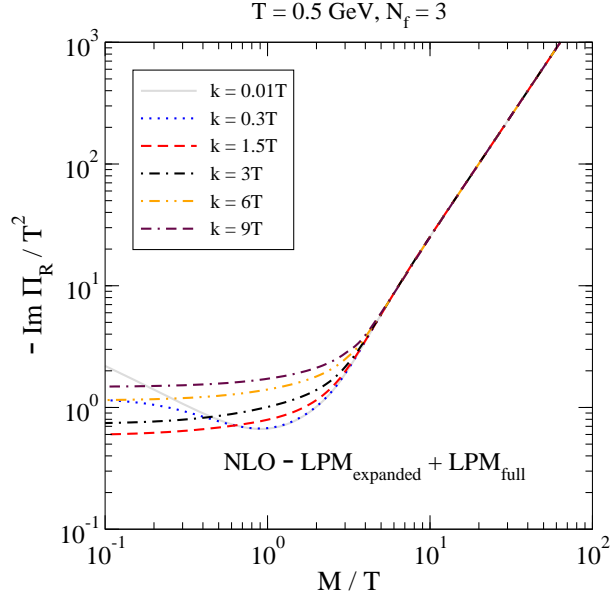}
}

 \caption[a]{The vector channel spectral function
 determined through \eq\nr{match} (from ref.~\cite{lpm}).
 }
 \label{fig:imNLO}
\end{figure}
%
%%%%%%%%%%%%%%%%%%%%%%%%%%%%%%%%%%%%%%%%%%%%%%%%%%%%%%%%%%%%%%%%%%%%%%%%%%%%

In \fig\ref{fig:screening}, 
screening masses measured 
in $\Nf = 2$ lattice QCD at $T \approx 250$~MeV are compared 
with corresponding predictions from perturbation theory, based on solving an 
LPM-type Schr\"odinger equation. It is seen that even at this ``low'' 
temperature, reachable in the current generation of LHC experiments, 
resummed perturbation theory does reproduce the qualitative features
of the lattice measurements, with differences 
only on the $\sim 15\%$ level. (It would be interesting to 
increase the resolution of the simulations, to take the continuum
limit, and to study several temperatures, in order to see if the 
remaining discrepancies decrease.) 

In \fig\ref{fig:imNLO} a result for $\im \Pi^{ }_\rmii{R}$
based on \eq\nr{match} is plotted
as a function of $M$ and $k$. A crossover from one type of behaviour 
at $M \lsim T$ to another shape at $M \gsim \pi T$ is clearly visible; 
as discussed above, the crossover takes parametrically
place at $M \sim g^{1/2} T$.

Finally, in \fig\ref{fig:pheno} the physical dilepton rate from 
\eqs\nr{match}, \nr{dilepton_rate} is illustrated. 
The grey bands indicated the 
uncertainty associated with variations of the renormalization scale. 
Even though the uncertainty is $\gsim 50\%$ for $M \lsim 1$~GeV, 
the overall shape of the curves as well as their general normalization
can be predicted. We also note that a recent study of the soft regime
up to NLO~\cite{gm} increases the results by 
$\rmO(10\%)$ for general parameters (up to 100\% for $M=0.25$~GeV), 
which suggests that the grey bands do indeed capture the 
magnitude of uncertainties.  

%%%%%%%%%%%%%%%%%%%%%%%%%%%%%% SECTION %%%%%%%%%%%%%%%%%%%%%%%%%%%%%%%%%%
%
\section{Outlook}
\label{se:outlook}

\begin{figure}[t]

\centerline{
    \includegraphics[angle=0,width=8.0cm]{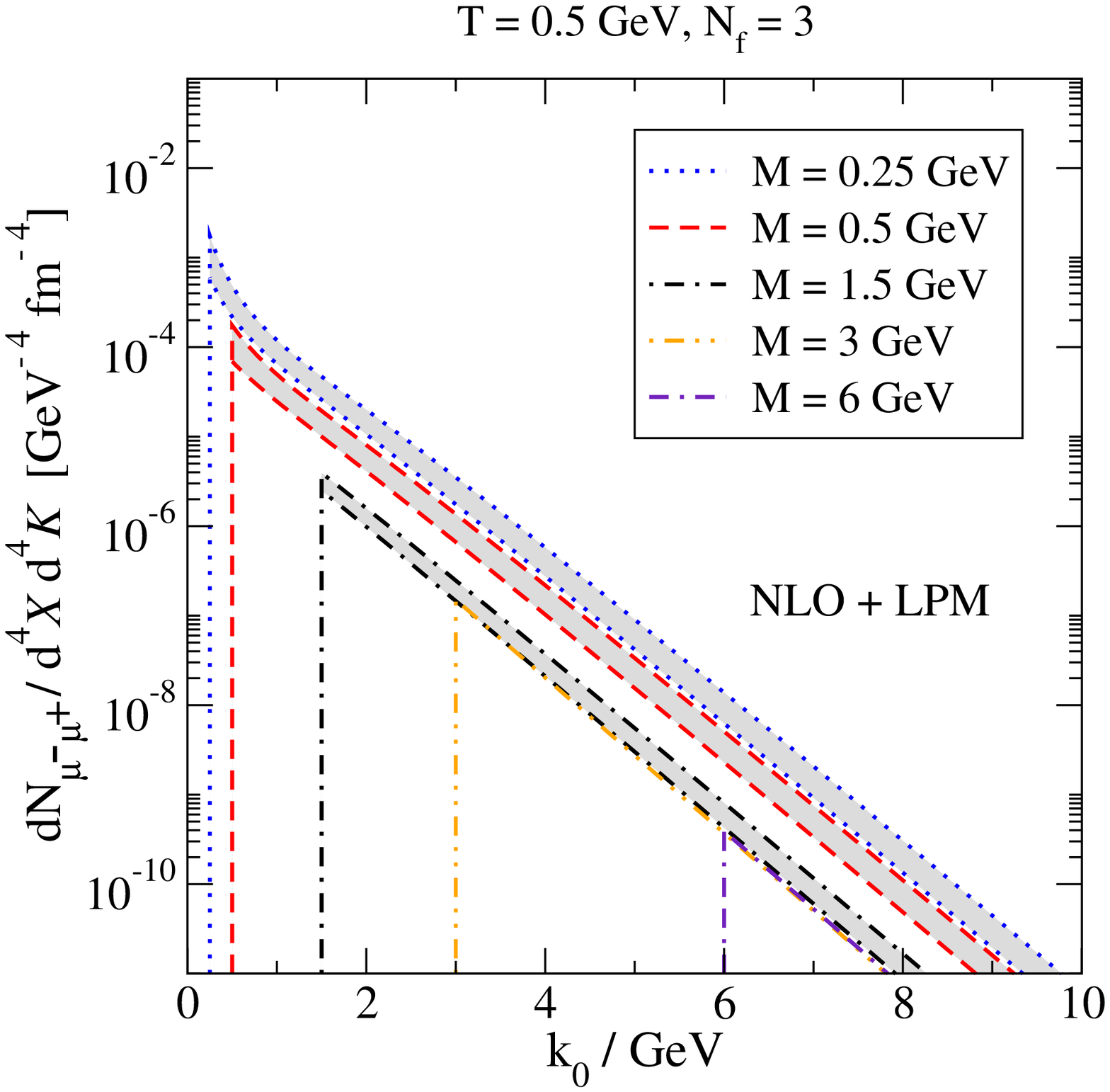}
}

 \caption[a]{Physical $\mu^-\mu^+$ production 
 rates based on \eqs\nr{match}, \nr{dilepton_rate}
 (from ref.~\cite{lpm}). 
 }
 \label{fig:pheno}
\end{figure}

There are two major ``applications'' for the results reviewed here.
First of all, curves such as shown in \fig\ref{fig:imNLO}, complemented
by results for the regime $\ko < k$, can be
inserted into \eq\nr{relation} and compared with direct lattice
measurements. Compilations suitable for this purpose, based on 
refs.~\cite{nlo3,lpm3,cond,dilepton,lpm,gm}, can be downloaded from 
ref.~\cite{dilepton-lattice}.
Second, results such as shown in \fig\ref{fig:pheno} could be 
inserted into hydrodynamical models such as ref.~\cite{hydro1}, 
and compared with experimental data from heavy ion collision
experiments. Data suitable for this purpose, from refs.~\cite{lpm,gm}, 
can be downloaded from ref.~\cite{dilepton-lpm}.
%

%%%%%%%%%%%%%%%%%%%%%%%%%%%% FIGURE %%%%%%%%%%%%%%%%%%%%%%%%%%%%%%%%%%%%%%%%%
%
%
%%%%%%%%%%%%%%%%%%%%%%%%%%%%%%%%%%%%%%%%%%%%%%%%%%%%%%%%%%%%%%%%%%%%%%%%%%%%



\bibliographystyle{elsarticle-num}
% \bibliography{hp13_biblio}

%% The provided style file elsarticle-num.bst formats references
%% in the required Procedia style


\end{document}